\documentclass[twocolumn,showpacs,preprintnumbers,amsmath,amssymb]{revtex4}

\usepackage{dcolumn}
\usepackage{bm}

\usepackage{amssymb}
\usepackage{epsfig}
\usepackage{graphicx}

\newcommand{\be}{\begin{equation}}
\newcommand{\ee}{\end{equation}}

\newcommand{\ba}{\begin{eqnarray}}
\newcommand{\ea}{\end{eqnarray}}

\newcommand{\n}[1]{\label{#1}}
\newcommand{\bea}{\begin{eqnarray}}
\newcommand{\eea}{\end{eqnarray}}

\newcommand{\hh}{\, ,\hspace{0.5cm}}
\newcommand{\hhh}{\, ,\hspace{0.2cm}}

\def\({\left(} \def\){\right)}

\newcommand{\bd}{\begin{displaymath}}
\newcommand{\ed}{\end{displaymath}}

\newcommand{\eq}[1]{(\ref{#1})}

\begin{document}

\title{A Toy Model for Topology Change Transitions: Role of Curvature
Corrections}

\author{Valeri P. Frolov}

\email{frolov@phys.ualberta.ca}

\affiliation{Theoretical Physics Institute,\\ University of Alberta,
Edmonton, Alberta, Canada T6G 2G7}

\author{Dan Gorbonos}

\email{gorbonos@phys.ualberta.ca}

\affiliation{Theoretical Physics Institute,\\ University of Alberta,
Edmonton, Alberta, Canada T6G 2G7}

\begin{abstract}
We consider properties of near-critical solutions describing a test
static axisymmetric $D$-dimensional brane interacting with a bulk
$N$-dimensional black hole $(N>D)$. We focus our attention on the
effects connected with curvature corrections to the brane action.
Namely, we demonstrate that the second order phase transition in
such a system is modified and becomes first order. We discuss
possible consequences of these results for merger transitions
between caged black holes and black strings.
\end{abstract}

  \ \ \

\maketitle

\section{Introduction}

Transitions with a change of Euclidean topology is a subject of wide
physical interest. One interesting example is the phase transition
connected with a nucleation of a black hole in a thermal bath.
Consider a thermal field with temperature $T$ in a flat spacetime.
One can use Euclidean fields on a spacetime with the topology
$R^{D-1}\times S^1$ to describe a canonical ensemble for such a
field. The size of the compact dimension $S^1$ is $\beta=1/T$. A
nucleation of a black hole changes the Euclidean topology from
$R^{D-1}\times S^1$ to $S^{D-2}\times R^2$. The corresponding
Euclidean space after the black hole nucleation is the
Gibbons-Hawking instanton~\cite{GH}.

Another important example of a similar phenomenon is the so-called
merger phase transition which occurs in models with large extra
dimensions when a black hole is localized in a spacetime which has
additional $k$ compact dimensions ($D=4+k$) (a caged black hole, for
reviews see~\cite{review,niarchos,reviewob1,reviewob2,reviewob3}).
In the absence of the black hole such a spacetime has the topology
$R^4\times T^k$. Kol argued~\cite{conifold-barak} that the black
hole-black string phase transition includes a local topology change
of the corresponding Euclidean manifold so that the singular
geometry is a cone over $S^{D-3}\times S^{2}$~\cite{endnote1}. This
topology change is similar to the conifold
transition~\cite{conifolds}. In the black hole phase the $S^{D-3}$
is contractible while in the black string phase the $S^{2}$ is
contractible. In order to achieve this topology change one has to
pass a configuration which is singular at the tip of the cone. The
``double-cone'' over $S^{D-3}\times S^{2}$ is given by \be\n{dc}
ds^{2}=d\rho^{2}+\frac{\rho^{2}}{D-2}\,\left[d\chi^{2}
+\cos(\chi)^{2}\,dt^{2}+(D-4)\,d\,\Omega_{D-3}^{2}\right]. \ee For
more details see~\cite{review,Barak2}.

Kol~\cite{Barak1} proposed that there exists a relation between
merger transitions and Choptuik's critical collapse
~\cite{choptuik-original,chptuik-review}. This correspondence can be
achieved by performing two analytic continuations. The physics of
the critical collapse and merger transitions have some common
features like a singular critical solution which turns out to be an
attractor and a self-similar solution in the neighborhood of the
singular point. A better understanding of one of the systems may
shed light on the other.

\begin{figure}[htb]
 \epsfxsize=40mm \epsfbox{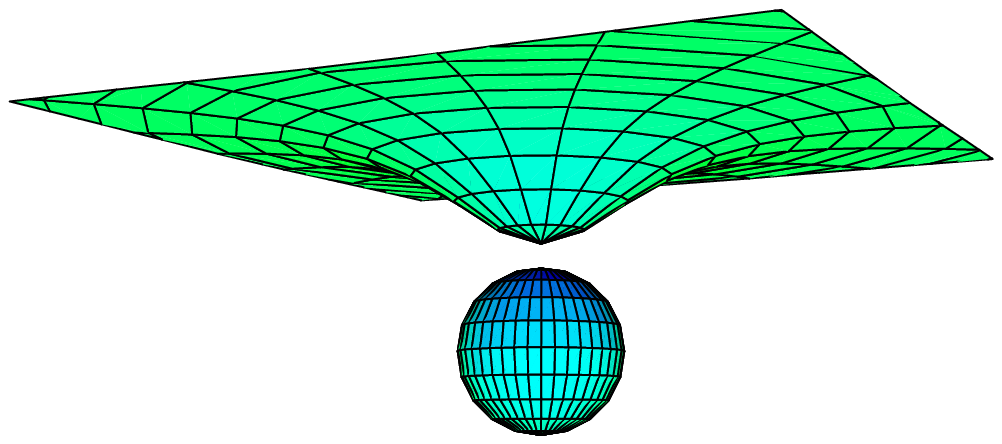}
 \epsfxsize=40mm \epsfbox{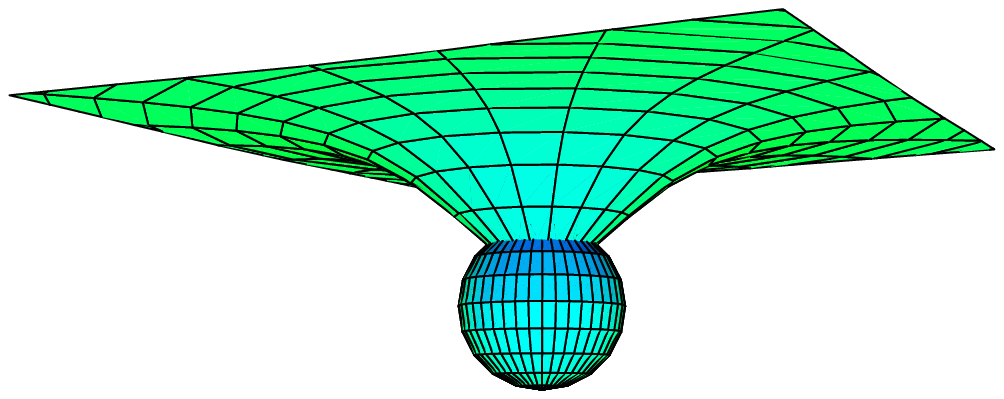}
\epsfxsize=40mm \epsfbox{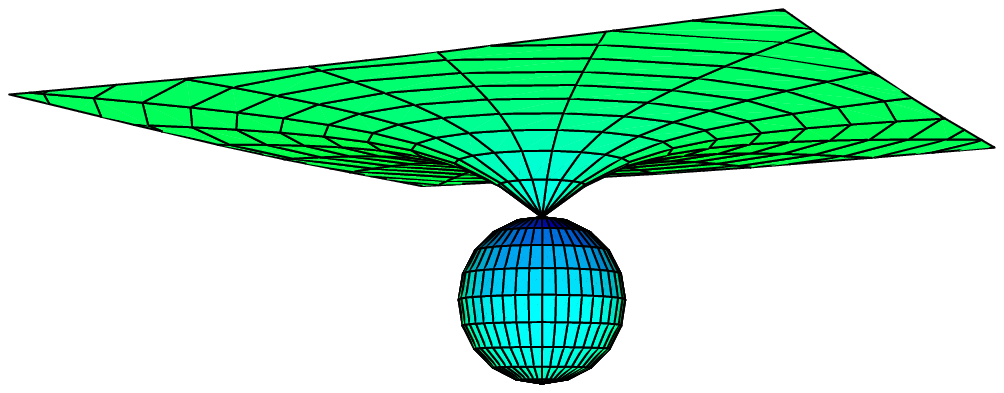} \caption{Three possible types
of configurations - the {\em subcritical} embedding (left) where the
brane does not touch the black hole horizon, the {\em supercritical}
(right) where we have an induced black hole on the brane (BBH), and
the {\em critical} (center) which is singular at its tip.}
\label{3con}
\end{figure}

It is interesting that there also exists a close similarity between
the properties of  {\em merger transitions} and  a toy model
proposed some time ago for study transitions during which the
Euclidean topology is changed~\cite{CFL,FLC,BBH1}. This model
consists of an $N$-dimensional static bulk black hole and a
$D$-dimensional brane ($D<N$) interacting with this black hole. The
brane is assumed to be a test brane and infinitely thin. The former
assumption means that one neglects the effects connected with the
gravitational field of the brane, while the latter one implies that
the effects of the brane thickness are neglected and its worldsheet
is a minimal surface which provides an extremum of the
Dirac-Nambu-Goto (DNG) action. It is assumed that the brane is
static and axisymmetric, so that the induced geometry on the brane
possesses the $O(D-1)$ group of isometry. It is also assumed that
far from the black hole the brane surface is parallel to the
equatorial plane of the bulk black hole. For such a brane there
exists two qualitatively different configurations: One, which is
called {\em subcritical}, is a brane which does not intersect the
black hole event horizon, and the other, {\em supercritical}, is a
brane crossing the horizon. In the latter configuration the induced
geometry on the world sheet of the brane is the geometry of
 a $D$-dimensional black hole, which is called a {\em brane black
hole}, or briefly {\em BBH}. Such a black hole is absent for a
subcritical configuration. Thus by changing the position of the
brane at infinity ({\em asymptotic data}), one generates a
transition between BBH and no-BBH phases (see Fig. \ref{3con}).

If this change is done adiabatically, then one deals with a one
parameter set of quasistatic solutions. After Wick's rotation of
time one gets a one parameter set of Euclidean induced metrics, with
a change of the Euclidean topology of the induced metric at some
critical value of the asymptotic data, which plays the role of an
order parameter. The Euclidean topology changes from $S^1 \times
R^{D-1}$ for the subcritical configuration to $R^2 \times S^{D-2}$
for the supercritical. It was demonstrated~\cite{CFL,FLC,BBH1} that
when the effects of the brane stiffness are neglected the relation
between the asymptotic data and the mass of the induced BBH for the
transition between sub- and supercritical configurations is {\em
universal}, that is it does not depend on the bulk black hole
characteristics. Moreover, there is no mass gap for a creation of a
BBH, so that the corresponding phase transition is of the second
order. The near-critical solutions possess discrete (for $D\le 6$)
or continuous (for $D>6$) self-similarity, which makes this
transition formally similar both to the merger transitions and the
near-critical collapse discovered by Choptuik~
\cite{choptuik-original}. (For a general review of the critical
collapse see e.g.~\cite{chptuik-review}. See also a
discussion~\cite{sorkin} of the critical collapse in a higher
dimensional spacetime.). These properties and close similarity of
near-critical solutions for both the BBH model and merger
transitions make it interesting to consider in the framework of the
BBH model some general problems which exist for this class of
models.

Before discussing these problems we mention that the universality of
the near-critical behavior in the BBH model is a consequence of the
following fact: only near-horizon properties of these solutions are
important . In this near-horizon domain there is no dimensionful
parameter which determines the behavior of the system. As a result
of this the system has scaling properties and the related phase
transition is of the second order. This is why the model captures
many universal features of various physical
systems~\cite{Filev:2008xt}. One important example of such a system
provides a holographic description of the meson melting phase
transition of matter in the fundamental
representation~\cite{Mateos:2007yp,Mateos:2007vn,Mateos:2006nu,Hoyos:2006gb,Babington:2003vm}.
The configuration consists of $N_c$ color Dp-branes and $N_f$ flavor
Dq-branes when $p<q$. The addition of the flavor Dq-branes is dual
to the addition of matter in the fundamental representation in the
gauge theory. In the limit $N_c \gg N_f$ the Dp-branes are described
by a p-brane supergravity action (``black Dp-branes'') while the
Dq-branes are described by the DNG action. In other words, we can
say that there are Dq-branes in the background of Dp-branes. In this
system the point, where the brane touches the black hole horizon
corresponds to a certain temperature where the mesons
melt~\cite{endnote2}. The holographic description of the melting
corresponds to the transition from the subcritical embedding to the
supercritical one.

Let us consider merger transitions in more detail. It is evident
that the double-cone solution \eq{dc} is smooth everywhere except
for the tip $\rho=0$ where the curvature becomes singular. Near the
tip the Kretchsman curvature scalar ${\cal
R}^2=R_{\mu\nu\rho\sigma}\,R^{\mu\nu\rho\sigma}$ infinitely grows
\be {\cal R}^2=\frac{4\,(D-3)^{2}\,(D-2)}{(D-4)\,\rho^{4}}\, . \ee
The existence of the infinite curvature indicates that the solution
obtained in the framework of classical Einstein gravity should be
modified by quantum corrections. In other words, the naked
singularity that is formed during the merger transition in its
classical description might be resolved by the inclusion of quantum
corrections into the classical action~\cite{endnote25}. This
conclusion is important. It means that if the transition between a
black hole and black string phases occurs through the merger
transition, one can expect the formation of a region with very high
(up to the Planckian) curvature in a system characterized by
macroscopic parameters (size of extra dimensions). An important
question is how quantum gravity effects modify an adopted picture of
classical merger transition.

Trying to answer this question one inevitably meets two difficult
problems. One is of technical origin, namely, how the near-critical
solutions for the merger problem are modified by quantum gravity
corrections, for example, by adding to the Einstein-Hilbert action
quadratic curvature corrections which arise in one loop
computations. One can expect that the corresponding corrections
become important when the curvature near the tip reaches the
Planckian scale. The other more difficult problem is the following.
At the corresponding Planckian scale the higher loop quantum gravity
terms might also become important. If this happens, it would
indicate that a complete solution of the problem requires the
summation of all quantum loops or the use of a more fundamental
theory of gravity, such as string theory. All of the above makes the
problem very complicated for analysis.

For this reason it is interesting to analyze a much simpler BBH
model which has qualitatively the same behavior as merger
transitions. Its critical solutions also have
 curvature singularity
at the cone tip where it touches the horizon. One can expect that
 adding terms quadratic in the extrinsic curvature to the classical DNG
Lagrangian, which are analogous to local one loop corrections in
quantum gravity, may ``cure'' this ``disease.'' Such curvature
corrections naturally arise as a result of the stiffness
effect~\cite{endnote3}. In the case of strings they were suggested
by Polyakov~\cite{polyakov}. We can think about such terms as
corrections that come from the finite thickness of the brane which
is ignored in the DNG action. The DNG action can be considered as
the zeroth order in the expansion in the width over a typical length
in the system~\cite{VilShel,CarGre}. Usually the small parameter in
such an expansion is the ratio of the thickness of the brane to the
characteristic radius of the brane bending. It is instructive first
to study effects connected with the leading order terms in this
expansion. In this analogy the thickness of the brane plays the role
similar to the Planck scale. Moreover, if one describes the brane as
a special topologically stable solution of some nonlinear field
theory, one may, in principle, answer not only the question of how
the quadratic curvature corrections modify the near-critical
solutions, but also investigate the complete field theoretical
object behavior in the near-critical regime. Effectively this
corresponds to the summation of all the stiffness corrections.

In the present paper we focus on the first problem, namely we will
analyze how the lowest order stiffness corrections modify the phase
transition in the BBH systems.

A good analogy that helps to understand the effect of stiffness
terms is a stiff bar. Consider the bending of a stiff bar. When we
take into account the effect of the stiffness of the bar, its
bending costs energy. Hence we cannot bend the bar as much as we
want and it would eventually break long before a sharp tip is
created. The sharp tip corresponds to the singular critical solution
of the DNG action. One might expect that inclusion of higher
derivative terms to the Lagrangian would prevent the creation of
such a singular solution and will form a first order phase
transition long before. Indeed, as we will see this is what happens
in the BBH system for the subcritical configuration.

The higher derivative corrections to the BBH system can serve as a
toy model for the singularity resolution of ``small BHs.'' Small BHs
are singular limits of BH parameters in which the horizon becomes
singular (see for example~\cite{Sen1,Sen2}). If we look at the
induced BH on the brane, the critical solution has
 a singularity exactly of this type.

The paper is organized as follows. In Sec.~{\it II} we review the
main results concerning the near-critical branes obtained in the
absence of stiffness in the BBH model. In Sec.~{\it III} we discuss
the curvature corrections for a stiff brane. The stiff brane
equations are presented in Sec.~{\it IV}. Section~{\it V} contains
the analysis of near-critical solutions in the linear approximation.
In Secs.~{\it VI} and~{\it VII} the numerical results for
near-critical branes are presented. Section~{\it VIII} contains a
summary of the obtained results and their discussion.

\section{Nonstiff Branes}

In this section we briefly review the main results of~\cite{BBH1}
concerning the behavior of near-critical $D$-dimensional branes
without stiffness interacting with a bulk static spherically
symmetric $N$-dimensional black hole. We do this mainly to explain
the set up of the problem and to fix the notations we will use
later. The Schwarzschild-Tangherlini metric of the bulk
$N$-dimensional spacetime is \be\n{met}
dS^2=g_{\mu\nu}dx^{\mu}dx^{\nu}=-F dT^2+F^{-1} dr^2 + r^2
d\Omega^2_{N-2}\, , \ee where $F=1-(r_g/r)^{N-3}$ and
$d\Omega^2_{N-2}$ is the metric of a $(N-2)$-dimensional unit sphere
$S^{N-2}$.  We define the coordinates $\theta_i$ ($i=1,\ldots,N-2$)
on this sphere by the relations \be d\Omega^2_{i+1}= d\theta^2_{i+1}
+\sin^2 \theta_{i+1} d\Omega^2_{i}\, . \ee

We denote by $x^\mu$ ($\mu=0,\ldots,N-1$)  the bulk spacetime
coordinates and by $\zeta^a\;(a=0,\ldots, D-1)$ the coordinates on
the brane worldsheet. The functions $x^{\mu}=X^{\mu}(\zeta^a)$
determine the brane world sheet describing the embedding of the
$(D-1)$-dimensional object (brane) in a bulk $N$-dimensional
spacetime. We assume that $D\le N-1$.

In the absence of stiffness, the brane configuration in an external
gravitational field $g_{\mu\nu}$ can be obtained by solving the
equations which follow from the DNG action~\cite{Dirac, Nambu,Goto}
\begin{equation}
S \: = \int d^D\zeta \sqrt{-\mbox{det}\gamma_{ab}}\, , \label{action}
\end{equation}
where $\gamma_{ab} \: = \: g_{\mu\nu}X^\mu_{,a} X^\nu_{,b}$ is an
induced metric on the brane world sheet. We set the brane tension
factor, which does not enter the brane equations, equal to 1. It is
well known that an extremum of this action is a minimal surface. Let
$n^{\mu}_{(i)}$ be unit normals to the brane, and \be
K^{(i)}_{\alpha\beta}=-\frac{\partial X^{\mu}}{\partial
\zeta^{\alpha}}\,\frac{\partial X^{\nu}}{\partial
\zeta^{\beta}}\,\nabla_{\nu}\,n^{(i)}_{\mu}\, , \label{ext_def}\ee
be an extrinsic curvature tensor. ($\nabla_{\nu}$ is a covariant
derivative with respect to the bulk metric $g_{\mu\nu}$.) Then the
nonstiff brane equations are of the form \be\n{breq}
K^{(i)}=g^{\alpha\beta}K^{(i)}_{\alpha\beta}=0\, . \ee

For the axially symmetric $D$-dimensional static brane (with the
isometry group $O(D-1)$)
 the induced metric is $(n=D-2$)
\be\n{mebr}
ds^2=\gamma_{ab}d\zeta^a d\zeta^b=
\ee
\[
-F dT^2+[F^{-1}+r^2 (d\theta/dr)^2] dr^2 + r^2 \sin^2\theta d\Omega^2_{n}\, ,
\]
and the action \eq{action} reduces to
\be\n{raction}
S=\Delta T {\cal A}_{n}\int dr {\cal L}\hhh
{\cal L}=r^{n}\, \sin^{n}\theta\, \sqrt{1+F r^2 (d\theta/dr)^2}\, .
\ee
Here $\Delta T$ is the interval of time, and ${\cal
A}_n=2\pi^{n/2}/\Gamma(n/2)$ is the surface area of a unit $n$-dimensional
sphere.

By analyzing the brane equation it is easy to show~\cite{BBH1} that
for a brane which asymptotically approaches the equatorial plane
$\theta=\pi/2$ one has \be\n{asq} \theta={\pi\over 2}+q(r)\hh
q={p\over r}+p'\left\{
\begin{array}{cc}
r^{-1}\ln r\, ,&\mbox{  for }n=1\, ,\\
r^{-n}\, ,&\mbox{  for }n>1\,.
\end{array}
\right. \ee We call the set of parameters $\{p,p'\}$, which
characterizes the solution, the {\em asymptotic data}.

The same solution can be determined by its behavior near the
horizon. A subcritical brane is uniquely specified by the distance
of its tip from the horizon. The condition of the brane surface
regularity at this point requires that its tangent plane at the tip
is orthogonal to the symmetry axis. This fixes the second constant
in the solution. Similarly, a regular brane crossing the horizon is
orthogonal to the horizon surface, so that a unique constant fixing
the solution is the ``gravitational'' radius of the induced BBH. A
solution  separating sub- and supercritical solution is a {\em
critical solution}. We denote by $\{p_*,p'_*\}$ its asymptotic data.

\begin{figure}[htb]
\begin{center}
 \epsfxsize=80mm
\epsfbox{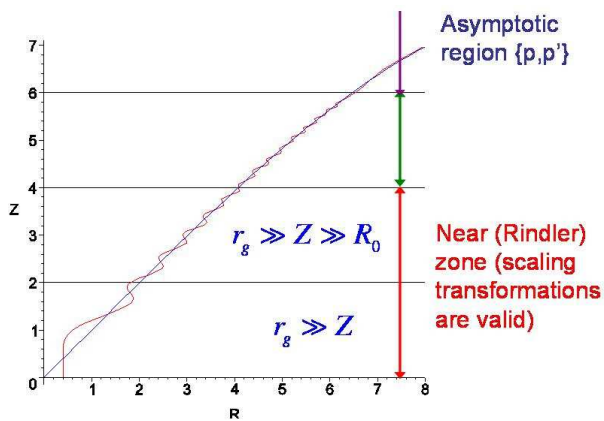}
 \caption{This figure schematically shows a configuration of a
supercritical brane in the regime when it is close to the critical
one. $Z$ is a proper distance from the horizon as a function of  the
radius $R$. $R_0$ is the radius of the surface of the intersection
of the brane with the horizon. In the region where $Z\ll r_g$ the
curvature surface of the horizon can  be neglected (the {\em Rindler
domain}).} \label{f1}
\end{center}
\end{figure}

Figure \ref{f1} illustrates the near-critical behavior of a
supercritical brane. (A similar graph for subcritical branes can be
easily obtained from this one by evident changes.) A near-critical
brane configuration is characterized by a parameter $R_0$, which is
its radius at the intersection with the horizon. For a subcritical
brane a similar parameter is $Z_0$, the proper distance of the tip
of the brane from the horizon. We consider a case when $R_0$ ($Z_0$)
is much smaller than the gravitational radius $r_g$ of the bulk
black hole. In the vicinity of the horizon, located at $Z=0$, that
is for $Z\ll r_g$, one has \be r-r_g\approx \kappa Z^2/2\hh
F\approx\kappa^2 Z^2\, , \ee where $\kappa={1\over 2}(dF/dr)|_{r_g}$
is the surface gravity. We call this region a {\em near (or Rindler)
zone}.  The corresponding induced metric for a near-critical brane
in the Rindler zone is \be\n{hmet} ds^2= -\kappa^2 Z^2 dT^2
+\left[\left({dZ\over d\lambda}\right)^2+\left({dR\over
d\lambda}\right)^2\right]d\lambda^2
 +R^2 d\Omega^2_{n}\, .
\ee
Here $(Z(\lambda),R(\lambda))$ is a brane equation written in a
parametric form. The action \eq{action} for this induced metric is
\ba
S&=&\kappa \Delta T {\cal A}_{n} {\cal S}\, ,\\
{\cal S}&=&\int d\lambda  Z R^n \sqrt{
(dZ/d\lambda)^2+(dR/d\lambda)^2}\, .\label{ex_action} \ea This
action is evidently invariant under the transformations
$\lambda\to\tilde{\lambda}(\lambda)$. In the regions where either
$Z$ or $R$ is a monotonic function of $\lambda$, these functions
themselves can be used as parameters. As a result, one obtains two
other forms of the action which are equivalent to ${\cal S}$ \be
{\cal S}=\int dZ {\cal L}_R=\int dR {\cal L}_Z\, , \ee where \be
{\cal L}_R=Z R^{n} \sqrt{1+{R'}^2}\hh {\cal L}_Z=Z R^{n}
\sqrt{1+{\dot{Z}}^2}\, . \ee Here the prime stands for the
derivative with respect to $Z$, while the dot stands for the
derivative with respect to $R$. The corresponding Euler-Lagrange
equations are \ba\n{eqR}
&&Z R R'' +(R R'-nZ)(1+{R'}^2)=0\, ,\\
&&R Z \ddot{Z} +(nZ \dot{Z}-R)(1+{\dot{Z}}^2)=0\, .\n{eqZ} \ea It is
easy to check that the form of Eqs. \eq{eqR}-\eq{eqZ} is invariant
under the following transformations: \ba\n{scR}
&&R(Z)=k\tilde{R}(\tilde{Z})\hh
Z=k\tilde{Z}\, ,\\
&&Z(R)=k\tilde{Z}(\tilde{R})\hh
R=k\tilde{R}\, .\n{scZ}
\ea

Equations \eq{eqR}-\eq{eqZ} have a simple solution \be\n{crit}
R=\sqrt{n} Z \ee which plays a special role. We call it a {\em
critical solution}. It describes a critical brane which touches the
horizon of the bulk black hole at one point, $Z=R=0$. It separates
the two different families of solutions, supercritical and
subcritical.

The relation between $R_0$ and $\{p,p'\}$ is of the form
\be
\ln R_0=\gamma \ln \Delta p +f(\ln \Delta p)+\ldots\,
\ee
where
\be
\gamma ={2\over n+2}\hh
\Delta p=\sqrt{(p-p_*)^2+(p'-p'_*)^2}\, ,
\ee
and the function $f(z)$ is
periodic, $f(z+\omega)=f(z)$, with the period
\be
\omega={\pi (n+2)\over \sqrt{4+4n-n^2}}\, .
\ee

Our aim is to study how the near-critical solutions are modified
when a brane is stiff. We assume that the effective width of the
brane is much smaller that the gravitational radius of the black
hole. In this approximation, as in the case of a nonstiff brane, the
main features of the phase transition in the BBH system are
determined by the brane behavior in the near zone, that is close to
the event horizon of the bulk black hole, where the Rindler
approximation is valid.

\section{Stiff Branes}

In this work we consider the Dirac-Nambu-Goto action with minimal
stiffness correction terms which play the role of higher curvature
corrections~\cite{Car}: \be S=-\int d^{n+2}\zeta
\sqrt{-\gamma}(1+B\,K^{2}+C\,{\cal K}^{2})\, . \label{theaction}\ee
Here $\gamma$ is the determinant of the induced metric given in Eq.
($\ref{action}$), $K=\sum_{i} K_{\mu}^{(i)\mu}$ is the trace of the
extrinsic curvature tensor, and ${\cal K}^{2}=\sum_{i}
K_{(i)\mu\nu}K^{(i)\mu\nu}$ is its square.  A minimal model of
stiffness corrections involves only quadratic powers of the
extrinsic curvature tensor. In this work we will concentrate on the
above ``truncated'' model as a toy model whose solution gives us the
static configuration of a stiff brane embedded close to the horizon
of the bulk black hole, namely, in the Rindler zone.

 In the particular
case of a domain wall the stiffness coefficients were calculated in
the framework of a microscopic model of a vacuum defect in field
theory with spontaneous symmetry breaking~\cite{CarGre}. In this
work we show that the exact numerical values of the coefficients do
not affect the qualitative features of the solution. Nevertheless
the sign of the coefficients is important. Positive stiffness
coefficients
\[ B,C>0
\] ensure us that the energy density for the static solution
\be  \epsilon=-{\cal L}=\sqrt{-\gamma}(1+B\,K^{2}+C\,{\cal K}^{2}),
\ee is positive since $K^{2}$ and ${\cal K}^{2}$ are both
non-negative.

Before going further let us discuss two interesting special cases of
the general theory.

{\bf $\bf C=0$ case.} In this special case the stiff string
equations have a simple exact solution. Namely, any solution of the
DNG equations \eq{breq} is at the same time a solution of the stiff
string equations. Indeed, a general variation of the action can be
split into the variation along the brane and the transverse one.
Variations along the brane surface vanish identically. For
transverse variations, multiplying the equations of motion by a
normal to the brane gives \bea n^{\mu}_{(i)}\,\frac{\delta {\cal
L}}{\delta X^{\mu}}=
K^{(i)}\,\(1+B\,K^{2}\) \nonumber\\
-2\,B\,\sqrt{-\gamma}\,K\,\,n^{\mu}_{(i)}\,\frac{\delta K}{\delta
X^{\mu}}=0. \n{eeqq}\eea Now, substituting the DNG equation
$K^{(i)}=0$ into the right hand-side of \eq{eeqq}, we see that it
vanishes. Hence $K^{(i)}=0$ is a solution of the stiff string
equations.

{\bf $\bf B+C=0$ case.} This case is not interesting for our
consideration since it violates the positive energy condition, but
it is of mathematical interest. Let us notice that the Gauss-Codazzi
relation for a flat bulk spacetime implies\be {\cal R}=K^{2}-{\cal
K}^{2}.\ee Thus one can rewrite the action~(\ref{theaction}) in the
following  form: \be S=-\int d^{n+2}\zeta \sqrt{-\gamma}(1+B\,{\cal
R}+(B+C)\,{\cal K}^{2})\, , \ee For $B+C=0$ the term with ${\cal
K}^{2}$ vanishes.

Let us return to the discussion of the general case. The units of
the stiffness coefficients are length-squared. Therefore under
scaling transformation of the spatial coordinates \be R\rightarrow
s\,R\;\;,\;\; Z\rightarrow s\,Z, \ee we have \be K^{2}\rightarrow
s^{-2}\,K^{2} \;\; ,\;\; {\cal K}^{2} \rightarrow s^{-2}\,{\cal
K}^{2}.\ee

Using this transformation we can set one of the coefficients to be
unit, say $C=1$. We can think about it as taking the basic unit
length of the stiff string to be $\sqrt{C}$. Then we are left only
with one free parameter $B$ in the action. We will use this choice
later in Secs.~$\ref{num1}$ and $\ref{num2}$, when we will discuss
the results of the numerical calculations.

For completeness we give here the components of the extrinsic
curvature for the brane $\(Z(\lambda),R(\lambda)\)$ with the induced
metric (\ref{hmet}):

\bea K_{\lambda\lambda}&=&{\cal
P}^{-1}{\cal A},\\
K_{TT}&=&-Z\,\frac{dR}{d\lambda}{\cal P}^{-1}, \nonumber\\
K_{\theta \theta}&=&R\,\frac{dZ}{d\lambda}{\cal P}^{-1},\nonumber
 \eea
 where
 \ba
 {\cal P}&=& \sqrt{
(dZ/d\lambda)^2+(dR/d\lambda)^2}\, ,\\
{\cal A}&=&\frac{dZ}{d\lambda}\,\frac{d^{2}R}{d\lambda^{2}}
-\frac{dR}{d\lambda}\,\frac{d^{2}Z}{d\lambda^{2}}\, .
\ea

The action (\ref{theaction}) for the induced metric~(\ref{hmet})
takes the following form: \ba
S&=&-\kappa \Delta T {\cal A}_{n} \int d\lambda {\cal L}\, \label{stiff_action}\\
{\cal L}&=&{\cal L}_0+B\,{\cal L}_1+C{\cal L}_2\, ,\nonumber\\
{\cal L}_0&=&Z\,R^n \,{\cal P}\, ,\nonumber\\
{\cal L}_1&=&Z\,R^n \,{\cal P}\,
\(\frac{{\cal A}}{{\cal P}^{3}}+\frac{dR/d\lambda}{Z\,{\cal P}}
+\frac{n\,dZ/d\lambda}{R\,{\cal P}}\)^{2},\nonumber\\
{\cal L}_2&=&{R^n \({dR/d\lambda}\)^2\over Z{\cal P}}+{n Z {\(d
Z/d\lambda\)}^2 R^{n-2}\over {\cal P}}\nonumber\\ &+&{Z R^n {\cal
A}^2\over {\cal P}^5}. \nonumber\ea The action (\ref{theaction}) is
evidently invariant under transformations
$\lambda\to\tilde{\lambda}(\lambda)$ as in the nonstiff case. In a
general case, a variation of this action gives equations containing
fourth derivatives, while the corresponding constraint equations are
of the third order~\cite{endnote4}.

\section{Euler-Lagrange equations for stiff branes}
  In the regions where either $Z$ or $R$ is a
monotonic function of $\lambda$, one of the coordinates can be used
as a parameter. As a result, one obtains two additional forms of the
action:

\be {\cal S}=-\int dZ {\cal L}_R=-\int dR {\cal L}_Z\, , \ee where
\bea {\cal L}_R=ZR^n {\cal P}\(1+B\left[\frac{R''}{{\cal
P}^{3}}+\frac{R'}{Z\,{\cal P}}+\frac{n}{R\,{\cal P}}\right]^2
\right. \nonumber
\\\left. +C\left[\frac{R''^{2}}{{\cal P}^{6}}+\frac{R'^{2}}{Z^{2}\,{\cal
P}^{2}}+\frac{n}{R^{2}\,{\cal P}^{2}}\right]\), \eea
\\
\bea {\cal L}_Z=ZR^n {\cal P}\(1+B[\frac{1}{Z\,{\cal
P}}-\frac{\ddot{Z}}{{\cal P}^{3}}+\frac{n\,\dot{Z}}{R\,{\cal
P}}]^{2}\right. \nonumber\\\left. +C[\frac{\ddot{Z}^{2}}{{\cal
P}^{6}}+\frac{1}{Z^{2}\,{\cal
P}^{2}}+\frac{n\,\dot{Z}^{2}}{R^{2}\,{\cal
P}^{2}}]\).\label{lagrangian_sub} \eea
${\cal P}$ in the first
relation means ${\cal P}=\sqrt{1+{R'}^2}$, while in the second
equation one has ${\cal P}=\sqrt{1+{\dot{Z}}^2}$.

As in the nonstiff case,
  the form with $R(Z)$ is useful for the description of the
supercritical brane while $Z(R)$ is more suitable for the
description of subcritical branes.

The equation for $R(Z)$ takes the following form:
\begin{widetext}
\be
-2\,b\,Z^{3}\,R^{3}\(1+R'^{2}\)^{2}\,R^{(4)}+4\,b\,Z^{2}\,R^{2}\,\(R'^{2}+1\)\left[5\,Z\,R\,R'\,R''-\(R'^{2}+1\)\(n\,Z\,R'+R\)\right]\,R^{(3)}
+F\(R,R',R'',Z\)=0, \label{Req}\ee

where \bea
&&F\(R,R',R'',Z\)=5\,b\,Z^{3}\,R^{3}\(1-6\,R'^{2}\)\,R''^{3}
+3\,b\,Z^{2}\,R^{2}\,\(1+R'^{2}\)\left[5\,R\,R'+n\,Z\,\(4\,R'^{2}-1\)\right]\,R''^{2}
\nonumber\\
&+&Z^{3}\,R^{3}\,\(1+R'^{2}\)^{3}\,R''
-Z\,R\,\(1+R'^{2}\)^{2}\(2\,\left[2\,b+3\,B\right]\,n\,Z\,R\,R'
+b\,R^{2}\,\left[R'^{2}-2\right]\right.\\
&+&\left.
n\,Z^{2}\,\left[b+3\,B-3\,B\,n+2\,b\,(n-2)\,R'^{2}\right]\)\,R''+\(1+R'^{2}\)^{3}\,\(R^{3}\,R'\left[Z^{2}+Z^{2}\,R'^{2}\right]-n\,Z\,R^{2}\,\left[Z^{2}+Z^{2}\,R'^{2}\right]\)
\nonumber\\
&-&\(1+R'^{2}\)^{3}\(\left[b-3\,B\,(n-1)\right]\,n\,Z^{2}\,R\,R'-b\,n\,Z\,R^{2}\,R'^{2}+b\,R^{3}\,R'\,\left[R'^{2}+2\right]+n\,[n-2]\,Z^{3}\left[b+B\,(n-1)+2\,b\,R'^{2}\right]\),
 \nonumber \eea
 and $b=B+C$.

A similar equation for $Z(R)$ is \be - 2\,b\,R^3\,{Z}^3\,{\left( 1 +
{\dot{Z}}^2
\right)}^2\,Z^{(4)}+4\,b\,R^{2}\,Z^{2}\,\(1+{\dot{Z}}^{2}\)\,\left[5\,R\,Z\,\dot{Z}\,\ddot{Z}-\(1+\dot{Z}^{2}\)\,\(n\,Z+R\,\dot{Z}\)\right]\,Z^{(3)}
+G\(Z,\dot{Z},\ddot{Z},R\)=0, \label{zr}\ee

where \bea &&
G\(Z,\dot{Z},\ddot{Z},R\)=5\,b\,R^{3}\,Z^{3}\,\(1-6\,\dot{Z}^{2}\)\,\ddot{Z}^{3}
+3\,b\,R^{2}\,Z^{2}\,\(1+\dot{Z}^{2}\)\left[5\,n\,Z\,\dot{Z}+R\,\(4\,\dot{Z}^{2}-1\)\right]\,\ddot{Z}^{2}
+R^{3}\,Z^{3}\,\(1+\dot{Z}^{2}\)^{3}\,\ddot{Z}\nonumber\\&-&R\,Z\,\(1+\dot{Z}^{2}\)^{2}\,\(2\,\left[3\,B+2\,b\right]\,n\,R\,Z\,\dot{Z}
-
b\,R^{2}\,\left[2\,\dot{Z}^{2}-1\right]+n\,Z^{2}\,\left[2\,b\,(n-2)+\(3\,B+b-3\,B\,n\)\right]\)\ddot{Z}
\nonumber\\
&-&R^{2}\,Z^{2}\(R-n\,Z\,\dot{Z}\)\(1+\dot{Z}^{2}\)^{4}
+\(1+\dot{Z}^{2}\)^{3}\(-b\,n\,R^{2}\,Z\,\dot{Z}+n\,R\,Z^{2}\left[b-3\,B(n-1)\right]\,\dot{Z}^{2}+b\,R^{3}\left[1+2\,\dot{Z}^{2}\right]\right.\\
&+&\left.
2\,b\,n\,(n-2)\,Z^{3}\,\dot{Z}+n\,(n-2)\,Z^{3}\,\dot{Z}^{3}\left[b+B(n-1)\right]\).\nonumber
 \eea
\end{widetext}

Let us denote \be l=\max\(\sqrt{B},\sqrt{C}\)\, . \ee $l$ has
dimensionality of the length. The extrinsic curvature corrections
are dominant for $R,Z\lesssim l$ where the stiff brane differs
significantly from the DNG brane. The significant effect of the
stiffness is localized in the region where the original DNG brane is
extremely bent. This happens in the neighborhood of the point
$R=Z=0$ which is defined by the length scale $l$. For $R,Z \gg l$
the solution for the stiff brane-BH system approaches the DNG
brane-BH system. In particular, $R=\sqrt{n}Z$ is the attractor
solution for the DNG brane-BH system and therefore it should be also
an attractor for the stiff brane-BH system.

\section{Near-Critical Modes}
\label{perturbations} Let us study linear perturbations to the
attractor solution for the case of a stiff brane. Our objective is
to obtain the modes in the neighborhood of the attractor far away
from the singular region $Z\gg l$ but still located in the Rindler
zone. The modes will guide us later in the setting of the boundary
conditions for stiff branes.

Let us substitute in the equation for $R$~(\ref{Req}) the following
expression: \be R(Z)=\sqrt{n}Z+\rho(Z), \ee and keep only linear
terms in $\rho(Z)$. Then we obtain the following linearized
equation: \bea
a_{0}^{s}(Z)\,\rho&+&a_\Omega{1}^{s}(Z)\,\rho'+a_{2}^{s}(Z)\,\rho''+a_{3}^{s}(Z)\,\rho^{(3)}+a_{4}^{s}(Z)\,\rho^{(4)} \nonumber\\
&=&2\,C\,(n-1)\,(n+1)^{2}\,n^{-\frac{1}{2}}Z,
\label{pert eq}\eea  \bea a_{0}^{s}(Z)&=&(n+1)\,\left[2\,B\(\,n-5\)-C(n+7)\right]+a_{0}(Z),\nonumber\\
a_{1}^{s}(Z)&=&(n+1)\,Z\left[2\,B\,\(16-n\)-17\,C\,(n-1)\right]+a_{1}(Z), \nonumber\\
a_{2}^{s}(Z)&=&-(n+1)\,Z^{2}\,\left[2\,B\,(n+1)+C(2\,n-1)\right]+a_{2}(Z), \nonumber\\
a_{3}^{s}(Z)&=&-4\,(B+C)(n+1)Z^{3},\nonumber\\
a_{4}^{s}(Z)&=&-2\,(B+C)Z^{4}. \eea Here $a_{i}(Z)$ are the
coefficients in the linearized DNG brane equations:
\bea a_{0}(Z)&=&(n+1)^{2}\,Z^{2},\nonumber\\
a_{1}(Z)&=&(n+1)^{2}\,Z^{3}, \nonumber\\
a_{2}(Z)&=&(n+1)\,Z^{4}.
 \eea
 Now let us take the limit $Z\gg l$ and as a result we obtain the following equation:
\bea
&&(n+1)^{2}\left[\,Z\,\rho+Z^{2}\,\rho'\right]+(n+1)\,Z^{3}\,\rho''\nonumber\\
&-&2\,(B+C)\,\left[2(n+1)\,Z^{2}\,\rho^{(3)}+Z^{3}\,\rho^{(4)}\right]
\nonumber\\&=&2\,C\,(n-1)\,(n+1)^{2}\,n^{-\frac{1}{2}}.\n{lin}
\eea

The leading term of the particular solution at large $Z$ is
\be\rho_{P}= \frac{C\,(n^{2}-1)}{\sqrt{n}\,Z}+{\cal
O}\(\frac{1}{Z^{2}}\),\ee and therefore it does not have an effect
on the attractor $R=\sqrt{n}Z$ at large $Z$, as expected. A general
solution of \eq{lin} is a sum of this particular solution and a
general solution of the homogeneous equation obtained from \eq{lin}
by omitting the right-hand side. It is interesting that this
homogeneous solution depends only on the sum $B+C$ of the stiffness
coefficients.

 Since the homogeneous equation is of the fourth order,
 it has four linearly independent solutions (asymptotic modes).
Two of the asymptotic modes reproduce the asymptotic solutions for
the DNG brane: \be \rho \sim
Z^{-\frac{1}{2}(n\pm\sqrt{n^{2}-4\,n-4})}. \label{asym}\ee

The other two modes appear only for the stiff brane:  \be \rho \sim
\exp\({\pm \sqrt{\frac{n+1}{2\,(B+C)}}\,Z}\). \ee The additional two
modes above are added due to the stiffness corrections. One of the
additional modes is an unstable mode which takes the solution away
from the attractor. This mode should be eliminated by appropriate
boundary conditions in order to reproduce the DNG solutions at large
distances.

Thus we arrive to a boundary value problem. Let us take, for
example, the subcritical configuration where the solution can be
written as $Z(R)$ (a similar discussion is applicable to the
supercritical configuration with some evident changes). Since
Eq.~(\ref{zr}) is of the fourth order, we need four initial values
(in case of an initial value problem). The configuration is axially
symmetric with the symmetry axis $R=0$ and it is plausible that the
stiff brane solution preserves the same axial symmetry. Consider a
brane passing through the point $Z(0)=Z_{0}$ (the proper distance of
the tip of the brane from the horizon). The axial symmetry and the
regularity of the brane at $R=0$ enforces $\dot{Z}(0)=0$ and
$Z^{(3)}(0)=0$, while $\ddot{Z}(0)$ remains a free parameter. This
free parameter will allow us to eliminate the unstable mode by
finding the specific value for $\ddot{Z}(0)$.

In conclusion, we have a boundary value problem defining
near-critical solutions of the stiff brane equations: In order to
obtain the right asymptotic behavior at large $R$
\[Z \rightarrow \frac{R}{\sqrt{n}}\]
for a brane which passes at $Z(0)=Z_{0}$ we have to tune the
parameter $\ddot{Z}(0)=\ddot{Z}_{0}$.

\section{Numerical Results for Stiff Branes: $n=1$ Case}
\label{num1} Using a numerical shooting analysis we obtained the
values of $\ddot{Z}_0$ for which there is a solution for the
subcritical brane that satisfies the boundary conditions: It starts
at $Z_0$ and asymptotically goes to the attractor. A similar
analysis was performed for the supercritical configuration where
values of $R''(0)$ were determined as a function of $R_{0}$ (the
radius of the BBH horizon).

\begin{figure} [htb]
\begin{center}
\epsfxsize=60mm \epsfbox{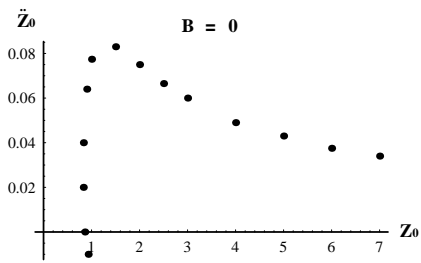} \caption{$\ddot{Z}_0$ as a
function of $Z_0$ for $n=1$ and $B=0$. The symmetry of the action in
this case implies that the same graph is valid to the supercritical
configuration as well -- $R''(0)$ as a function of $R_0$.}
\label{figB0}
\end{center}
\end{figure}

Let us start by examining the case $n=1$. As we will see, this case
is qualitatively different from $n>1$. For $B=0$ the action
[Eq.~(\ref{stiff_action})] is completely symmetric for the
interchange of $R\leftrightarrow Z$. This discrete symmetry of the
equations implies the same results for the subcritical and
supercritical configurations. For this reason we give here the
results for the subcritical configuration $B=0$ and $n=1$
(Fig.~\ref{figB0}) when for the supercritical configuration the
graph is the same (up to the interchange of $R\leftrightarrow Z$).
The plot at this figure shows $\ddot{Z}_0$ as a function of $Z_0$
for near-critical configurations. The finite gap in the neighborhood
of the point $R=Z=0$ demonstrates the first order phase transition
in the BBH system. A detailed interpretation of this picture is
given below in the discussion of the case $n=2$ (where the same
picture appears only in the subcritical configuration).

For the case of $B \neq 0$ the action (\ref{stiff_action}) is no
longer symmetric under the reflection
 $R\leftrightarrow Z$. Nevertheless we find
numerically that the results are symmetric within the used accuracy.
Evidence for this symmetry we can find in the linearized equations
for a perturbation around the attractor. In Sec.~\ref{perturbations}
we studied linearized perturbations to the supercritical
configuration \be R(Z)=\sqrt{n}Z+\rho(Z). \ee Keeping only linear
terms in $\rho(Z)$ gave us Eq. ($\ref{pert eq}$). For the special
case of $n=1$ this equation reads \bea &&
(B+C)\,Z^{4}\,\rho^{(4)}(Z)+
4\,(B+C)\,Z^{3}\,\rho^{(3)}(Z)\nonumber\\&+&Z^{2}\(4\,B+C-Z^{2}\)\,\rho''(Z)-2\,Z^{3}\,\rho'(Z)\nonumber
\\&-&2\(Z^{2}-4\,B-4\,C\)\,\rho(Z)=0 \eea

In a similar way for the subcritical configuration substitution of
\be Z(R)=\frac{R}{\sqrt{n}}+\zeta(R) \ee into Eq. (\ref{zr}) and
keeping only linear terms in $\zeta(R)$ gives the following linear
equation ($n=1$)
 \bea
 &&(B+C)\,R^{4}\,\zeta^{(4)}(R)+4\,(B+C)\,R^{3}\,\zeta^{(3)}(R)\nonumber\\
&+&R^{2}\(4\,B+C-R^{2}\)\,\ddot{\zeta}(R)-2\,R^{3}\,\dot{\zeta}(R)\nonumber \\
&-&2\(R^{2}-4\,B-4\,C\)\,\zeta(R)=0.
 \eea

Hence the symmetry $R\leftrightarrow Z$ is demonstrated analytically
in the linear approximation. This does not imply an exact reflection
symmetry of the solutions, but at least makes it possible.

\section{Numerical Results for Stiff Branes: $n>1$ Case}
\label{num2} For $n>1$ there is no reflection symmetry of the action
anymore, and sub- and supercritical solutions behave quite
differently. Let us start with the subcritical configuration and
demonstrate that it exhibits the same qualitative features as $n=1$.
This is a good place to compare the details of the new picture with
the nonstiff case.

\begin{figure}[htb]
\epsfysize=50mm \epsfbox{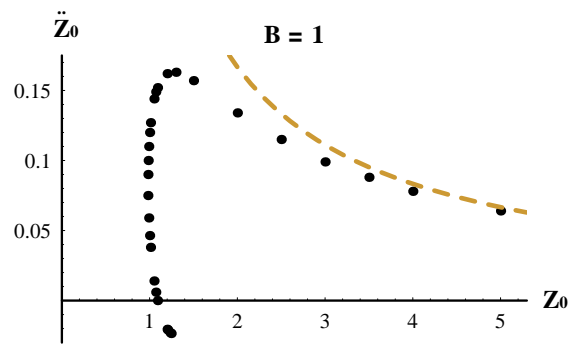} \caption{$\ddot{Z}_0$ as a
function of $Z_0$ for $n=2$. The dashed line is the same function
for DNG branes (without stiffness terms).} \label{B1n2}
\end{figure}

Consider, for example, the case of $n=2$ and $B=1$ in
Fig.~\ref{B1n2}. The value of $\ddot{Z}_0$ is plotted as a function
of the position where the brane crosses the axis of symmetry $Z_0$.
In the case of DNG branes, i.e. without stiffness, the dependence of
$\ddot{Z}_0$ on $Z_0$ is determined from the Euler-Lagrange equation
(\ref{eqZ}) to be (see in~\cite{BBH1}): \be
\ddot{Z}_0=\frac{1}{(n+1)\,Z_0}. \label{init_DNG}\ee This function
is plotted in Fig.~
\ref{B1n2} for comparison with the case of stiff
branes.

Few features can be observed in the graph:
\begin{itemize}
\item There is a finite gap $0<Z_0\lesssim 1$ in which the solution for the embedded stiff brane does not exist at
all. Hence the singular point is resolved for the subcritical
branch. This is a characteristic feature of first order phase
transitions.
\item $\ddot{Z}_0$ is bounded, unlike DNG branes [Eq.~
(\ref{init_DNG})]
for which $\ddot{Z}_0$ is unbounded
\item For $1 \lesssim Z_0 \lesssim 1.25$ we see coexistence of two
branches of solutions. For any $Z_0$ in this range there are two
possible values of $\ddot{Z}_0$ and thus two possible configurations
of the stiff brane. One can compare the energy [see
\eq{lagrangian_sub}] of the two branches.  A numerical comparison of
the energies [Fig.~\ref{energy fig}] reveals that the branch with
the lower values of $\ddot{Z}_0$ is energetically favored. The
branch with higher energy corresponds to a local phase at maximum.
This solution should be unstable and separates two stable phases in
a first order phase transition.
\begin{figure}[htb]
\epsfxsize=60mm \epsfbox{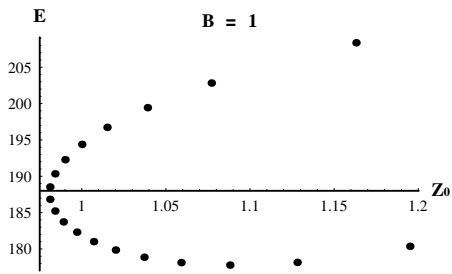} \caption{The energy density
integrated for $0\leq R \leq5$ as a function of $Z_0$ comparing two
branches in the segment ($1 \lesssim Z_0 \lesssim 1.25$). Note that
the minimal energy is obtained at the point which corresponds
approximately to $\ddot{Z}_0=0$.} \label{energy fig}
\end{figure}

\item There are solutions for stiff branes that satisfy the boundary conditions with negative $\ddot{Z}_0$. Such
solutions exist only for $-0.025\lesssim \ddot{Z}_0 \lesssim 0$.

\item There exists an ``end point'' in the plot with minimal value
of $\ddot{Z}_0$. For $\ddot{Z}_0$ less than this value a solution
does not exist.

\item At large values of $Z_0$ we see that the effects of stiffness are
negligible. The points of the stiff branes approach the DNG branes
at large values of $Z_0$.
\end{itemize}

In order to check that the obtained results are robust we repeated
the same calculations for various values of $B$. In all cases we
found that the same qualitative behavior repeats itself: A finite
gap in the existence of solutions for $0<Z_0\lesssim l$,
 two branches of solutions in a small neighborhood of $Z_0 \sim
 l$ etc.

\begin{figure}[htb]
\epsfxsize=55mm \epsfbox{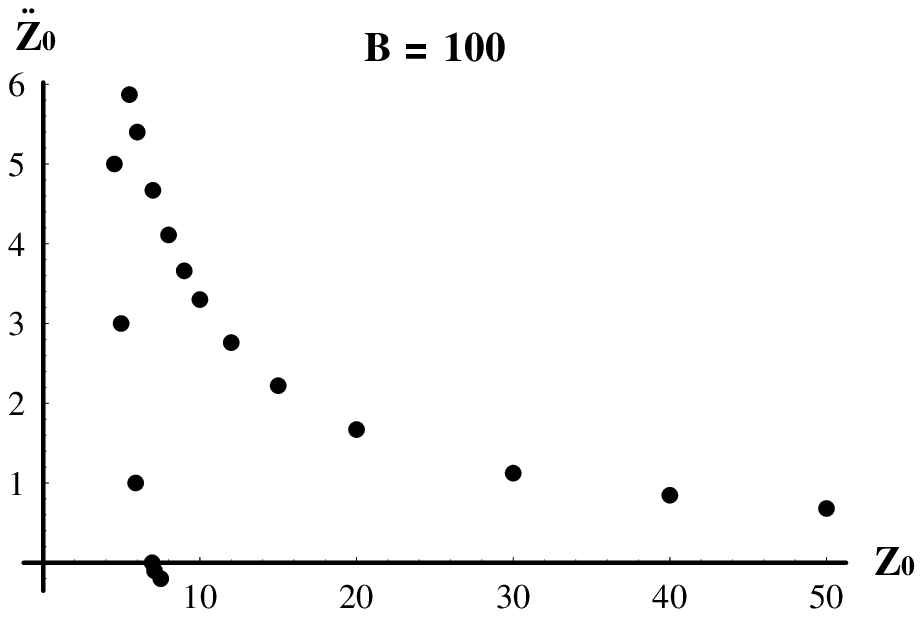} \epsfxsize=60mm
\epsfbox{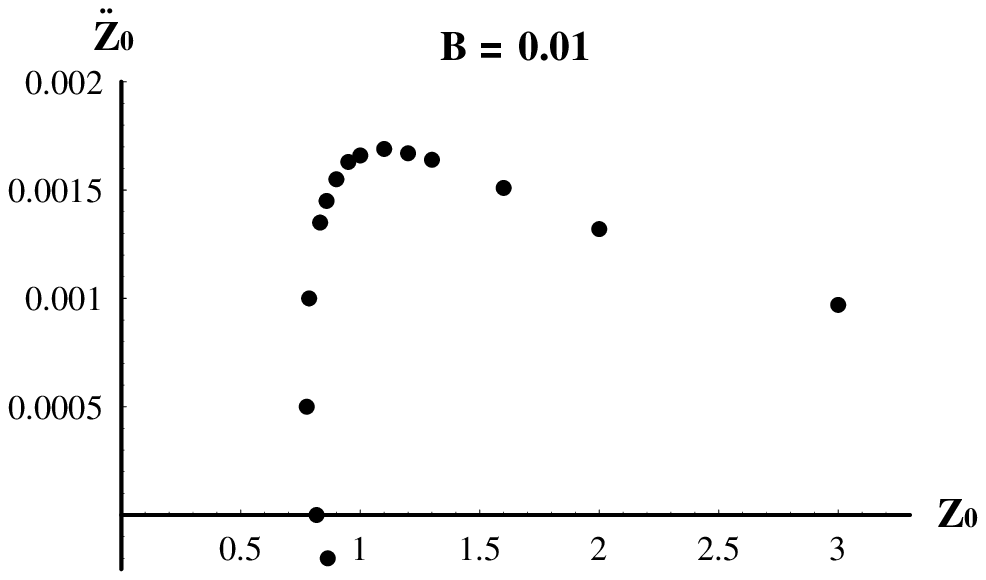}
 \caption{
$\ddot{Z}_0$ as a function of $Z_0$ for $n=2$. $B=100,\; \;
0.01$.}\label{two plots}
\end{figure}

For illustration we give in Fig.~\ref{two plots} two graphs for two
values of $B$ with four orders of magnitude difference
($B=0.01,\;100$).

\begin{figure}[htb]
\epsfxsize=60mm \epsfbox{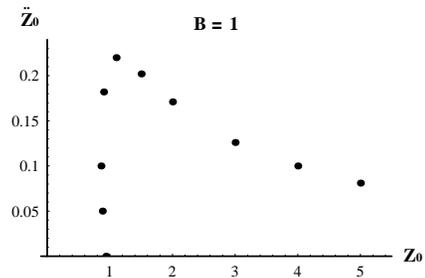}
 \caption{
$\ddot{Z}_0$ as a function of $Z_0$ for $n=4,\, B=1$.} \label{n=4}
\end{figure}

In addition we checked for various dimensions $n=4,5$ and found the
same qualitative behavior (see Fig.~\ref{n=4} for $n=4$ as an
example). Despite the fact that $n \leq 4$ is different from $n \geq
5$ since in the former the phase space behavior of the critical
solution behaves as of a focal point and in the latter as a node
(see~\cite{BBH1}). This type of transition in the near-critical
solutions has no influence on the neighborhood of the singular point
$R=Z=0$ where the stiffness terms are dominant.

\begin{figure}[htb]
\epsfxsize=60mm \epsfbox{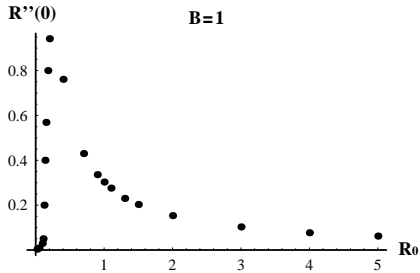}
 \caption{
$R''(0)$ as a function of $R_0$ (supercritical) for $n=2,\quad
B=1$.} \label{superlow}
\end{figure}

\begin{figure}[htb]
\epsfxsize=60mm \epsfbox{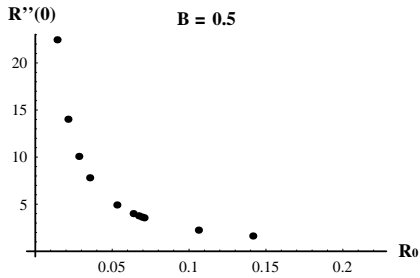}
 \caption{
$R''(0)$ as a function of $R_0$ (supercritical) for $n=2,\,\quad
B=0.5$.} \label{superhigh}
\end{figure}

It is surprising that when we repeated similar calculations for {\em
supercritical} stiff branes with $n>1$ we found that for the
supercritical configurations  there is no singularity resolution.
The stiffness terms break the symmetry between the supercritical and
subcritical brane-black hole systems. The supercritical solutions
show no gap nor double-branch behavior. As an example let us look at
the supercritical configuration for $n=2$. For $B<0.906$ we did not
find evidence for the existence of a solution in the vicinity of the
point $R(0)=R''(0)=0$. See Fig.~\ref{superlow} for $B=1$ and
Fig.~\ref{superhigh} for $B=0.5$. We stress that in both cases the
curvature singularity still exists.

\noindent

\section{Discussion}

We observed that due to the stiffness corrections the singularity of
the critical solution is resolved for $n=1$ in a symmetric form
(both in the subcritical and supercritical configurations) and for
$n>1$ only for subcritical configurations. We observed this
resolution in the creation of a finite gap and a clear signature of
a first order phase transition. This signature is observed in a
typical hysteresis curve of coexistence of two phases--stable and
unstable. For $n=1$ we see a first order phase transition on both
sides of the singularity (supercritical and subcritical
configurations) when for $n>1$ we see only half of this picture--a
first order phase transition in the subcritical configuration.

We expect that a similar picture would emerge in merger transitions
when higher derivative corrections are included. Inclusion of higher
derivative corrections might cause the merger transition to become
 first order in nature and create a finite gap between the thin
black string (``the waist'') and the caged black hole. This way the
naked singularity and the violation of cosmic censorship hypothesis
that appear in the classical approximation would be resolved. This
might also be a natural way to resolve the apparent tension between
the suggested scenario for the merger transition and the observation
that such a pinch-off can occur only at infinite affine parameter
along the horizon~\cite{Horowitz:2001cz}. The resolution is the
following. When the system approaches the Planckian scale, at a
finite time, the first order phase transition takes the system to
the second phase. Therefore with quantum corrections the
``pathologies'' of infinite affine parameter and naked singularity
would be resolved.

The asymmetry that we found might be a result of the incompleteness
of the truncated model that we used to describe the full effect of
quantum corrections. It might be also a hint on an asymmetry which
is generic in the topology change in general.\\

\section*{Acknowledgments}    One of the authors (V.F.) is
grateful to the Research Program on Gravity and Cosmology at the
Yukawa Institute where this work was started. He also thanks the
Natural Sciences and Engineering Research Council of Canada (NSERC)
and the Killam Trust for financial support. DG thanks NSERC for the
financial support, and thanks Vadim Asnin, Amit Giveon, Barak Kol,
Matthew Lippert and Amos Ori for fruitful discussions.


\begin{thebibliography}{}
\bibitem{endnote1}
The $S^{2}$ is the time circle fibred over a segment in the $r-z$
plane when $r$ is the radial coordinate and $z$ is the coordinate of
the compact dimension
\bibitem{endnote2}
Higher derivative corrections to this configuration correspond to
finite 't Hooft coupling corrections in the gauge theory. In our
approximation of Rindler spacetime the higher derivative corrections
to the bulk BH (or the Dp-brane) are suppressed and therefore we
consider only corrections to the brane (or the Dq-brane). Therefore
this work might give guidelines for the necessary modifications to
the holographic picture of meson melting.
\bibitem{endnote3}
Inclusion of linear order corrections in the extrinsic curvature
will depend on a choice of sign and thus on an orientation
convention and we restrict ourselves to lagrangians that depend only
on the embedding of the worldsheet. See~\cite{Car} for a discussion
of this point.
\end{thebibliography}
\end{document}